\documentclass[proceedings]{JHEP37}
\usepackage{eurosym}
\usepackage{amssymb}
\usepackage{amsfonts}
\usepackage{amsmath}
\usepackage{epsfig}

\newbox\mybox
\makeatletter 
\def\lambdabar{\protect\@lambdabar}
\def\@lambdabar{\relax
\bgroup
\def\@tempa{\hbox{\raise.73\ht0
\hbox to0pt{\kern.25\wd0\vrule width.5\wd0
height.1pt depth.1pt\hss}\box0}}\mathchoice{\setbox0\hbox{$\displaystyle\lambda$}\@tempa}{\setbox0\hbox{$\textstyle\lambda$}\@tempa}{\setbox0\hbox{$\scriptstyle\lambda$}\@tempa}{\setbox0\hbox{$\scriptscriptstyle\lambda$}\@tempa}\egroup
}
\makeatother

\newcommand\fverb{\setbox\mybox=\hbox\bgroup\verb}
\newcommand\fverbdo{\egroup\medskip\noindent\fbox{\unhbox\mybox}\ }
\newcommand\fverbit{\egroup\item[\fbox{\unhbox\mybox}]}
\conference{Quantum, NC and MOND corrections to the entropic law of gravitation}
\abstract{Quantum and noncommutative corrections to the Newtonian law of inertia are considered in the general setting of Verlinde's entropic force postulate. 
We demonstrate that the form for the modified Newtonian dynamics (MOND) emerges in a classical setting by seeking appropriate corrections in the entropy.
We estimate the correction term by using concrete coherent states in the standard and generalized versions of Heisenberg's 
uncertainty principle. Using Jackiw's direct and analytic method we compute the explicit wavefunctions for these states producing minimal length as well as minimal products. Subsequently we derive a further selection criterium restricting the free parameters in the model in
providing a canonical formulation of the quantum corrected Newtonian law by setting  up the Lagrangian and Hamiltonian for the system.}

\title{Quantum, noncommutative and MOND corrections to the entropic law of
gravitation}
\author{Bijan Bagchi$^\circ$ and Andreas Fring$^\bullet$ \\
$\circ$ Department of Physics, Shiv Nadar University, Dadri, UP 201314, India%
\\
$\bullet$ Department of Mathematics, City, University of London,\\
$\,\,$ Northampton Square, London EC1V 0HB, UK \\
E-mail: bbagchi123@gmail.com, a.fring@city.ac.uk}

\input{tcilatex}
\begin{document}

\section{Introduction}

The weak equivalence principle is a well known concept, see e.g. \cite%
{berry1989}, that identifies the inertial mass $m_{I}$ occurring in Newton's
second law of motion $\vec{F}=m_{I}\vec{a}$, with the gravitational mass $%
m_{G}$ in Newton's inverse square law of gravitation. The latter accounts
for the attractive force between a body of mass $m_{G}$ at the position $%
\vec{r}$ and $n$ different others specified by their masses $m_{i}$
occupying positions ${\vec{r}_{i}}$, $i=1,2,...,n$, as

\begin{equation}
\vec{F}=-\sum_{i}\frac{Gm_{G}m_{i}(\vec{r}-\vec{r_{i}})}{|\vec{r}-\vec{r_{i}}%
|^{3}},\quad i=1,2,...,n,  \label{1}
\end{equation}%
where $G$ is the gravitational constant. Equating these two expressions for
the force when using $m_{I}=m_{G}$ readily yields an expression for the
acceleration of a particle in a gravitational field

\begin{equation}
\vec{a}=-G\sum_{i}\frac{m_{i}(\vec{r}-\vec{r_{i}})}{|\vec{r}-\vec{r_{i}}|^{3}%
},\quad i=1,2,...,n.  \label{2}
\end{equation}%
A consequence of this is the curious feature, known since the time of
Galileo, that objects that are dropped from some height, say from the top of
a building, will arrive at the same time on the ground as long as their
motion is not affected by air resistance or other disturbances, i.e. they
fall at the same rate with equal accelerations.

By invoking the holographic principle in the vicinity of a black hole E.
Verlinde \cite{verlinde2011origin,verlinde2017emergent} demonstrated
recently that Newton's second law of motion for a particle when confronted
with the law of gravitation for celestial bodies, can be viewed as entropic
in character. Employing the holographic argument that any information of the
black hole, which is imagined as a sphere of Schwarzschild radius $R$, can
emerge only from its surface (because what is inside the black hole remains
totally intractable to an outside observer), and using the equipartition
rule gave strong thermodynamical evidence to justify such a claim. Note that
the boundary of the black hole sphere which is basically an equipotential
surface acts as a holographic screen also popularly referred to as the event
horizon.

Assuming the change of entropy $\Delta S$ near the holographic screen to be
linear in the displacement $\Delta x$ of a test particle, Verlinde suggested
for $\Delta S$ the relation

\begin{equation}
\Delta S=2\pi k_{B}\frac{mc}{\hbar }{\Delta x}=2\pi k_{B}\frac{\Delta x}{%
\lambdabar%
},  \label{V1}
\end{equation}%
with $%
\lambdabar%
:=\hbar /mc$ denoting the reduced Compton wavelength, $k_{B}$ the Boltzmann
constant, $\hbar $ the reduced Planck constant and $c$ the speed of light.
The change in entropy was also assumed to generate an entropic force $F$
acting on the particle to be of the form

\begin{equation}
F\Delta x=T\Delta S,  \label{V2}
\end{equation}%
where $T$ is the temperature. Taking $T$ to be given by Unruh's temperature 
\cite{unruh1976notes} for an accelerated observer, namely

\begin{equation}
k_{B}T=\frac{\hbar }{2\pi }\frac{a}{c},  \label{V3}
\end{equation}%
where $a$ stands for the acceleration of the particle, consistency gave
Newton's formula for the second law of motion $F=ma$ when combining (\ref{V1}%
)-(\ref{V3}). Next, to arrive at the law of gravitation by restricting to
the spherical boundary having an area $A=4\pi R^{2}$, $R$ being the radius
of the sphere, he made use of the holographic principle that the total
number of bits making up the maximally storage space is proportional to $A$.
This gave the number $N$ of used bits as $N=Ac^{3}/G\hbar $, $G$ being
identified as Newton's gravitational constant as in (\ref{1}). Using the
equipartition rule for the average energy for every bit, $E=\frac{1}{2}%
Nk_{B}T=mc^{2}$, $M$ denoting the mass in the part of the space enclosed by
the holographic screen, yielded the well known Newton's law of gravitation: $%
F=-GMm/R^{2}$.

It is worthwhile to recall some history behind Verlinde's formulation.
First, an early work by Jacobson \cite{jacobson1995} attempted to derive
Einstein's relativistic equations from pure thermodynamical considerations
by making the constant of proportionality between the area and entropy
universal following a preceding work of Bekenstein \cite{bekenstein1973} who
in turn looked at the entropy of any isolated system to be bounded by its
area. Second, Padmanabhan \cite{padmanabhan2004,padmanabhan2010} arrived at
a result of gravitational acceleration by reversing Unruh's
temperature-acceleration relation. Both Bekenstein as well as Padmanabhan's
discussions were carried out in a fully relativistic frameworks which
apparently have no analogue in Verlinde's non-relativistic formulation.

An important observation made by Verlinde was that since the maximally
allowed information stored in any continuum volume of space can only be
finite, it is not sensible to talk of localizing a particle with an infinite
degree of accuracy. Even though in the end he obtained the classical results
which were devoid from the appearance of $\hbar $, it is pertinent to bear
in mind that while the individual expressions of the change in entropy as
well as Unruh's temperature contain an explicit presence of ${\hbar }$, the
latter fortunately cancels out when we look for a force-acceleration
relationship. A question then naturally arises as to what happens if we seek
higher corrections to the uncertainty principle as is needed to accommodate
various modifications of the short distance structure in quantum theories
that attempt to incorporate gravity \cite{Kempf2}. This in turn calls for an
introduction of a so-called minimal length beyond which a localization of
space-time events is no longer possible. With a minimal observable length $%
\Delta x=0$ that is characteristic of a physical quantum state, it is
evident that an eigenstate with a zero-uncertainty in position can no longer
depict a physical state. Models in string theory \cite{String1,String2} as
well as in quantum gravity \cite{Rovelli} do indeed support the existence of
such a minimal length \cite{Hossenfelder}. Alternatively one may also view $%
\Delta x$ as a change in the black hole radius \cite{xiao2010}.

Our manuscript is organized as follows: In section 2 we discuss how
Verlinde's argument can be modified by including quantum or noncommutative
corrections in the energy and/or the entropy. We compute the correction
terms to the gravitational force for various choices of the free parameters
in the standard approach and generalized Heisenberg's uncertainty relations.
We use concrete expressions for the uncertainties obtained from different
types of coherent states whose wave functions we derive explicitly in
section 3 using Jackiw's direct and analytic method. In section 4 we derive
some Lagrangians and Hamiltonians for the corrected entropic force, which
turn out to be explicitly time-dependent. Demanding that the damping to be
small provides a further criterion that allows to exclude certain choices of
the free parameters. We state our conclusions in section 5.

\section{\noindent Quantum corrections to Newton's second law}

Let us now see how the above effects might be incorporated into the above
reasoning by modifying the equations (\ref{V1})-(\ref{V3}) and exploiting
Verlinde's observation that a strict localization of the test particle is
not possible. As argued by Santos and Vancea \cite{santos2012} the total
energy also depends on the momentum $p$ in form of the kinetic energy or
possibly in a more general way. This means that the uncertainty in the total
energy $\delta E$ could also acquire a term that depends on the uncertainty
in the momentum $\delta p$. We assume here the form%
\begin{equation}
\delta E=F\delta x+\alpha \frac{p}{m}\delta p=T\delta S,  \label{E1}
\end{equation}%
to be valid at thermal equilibrium where $\alpha $ is dimensionless,
possibly a constant. Note that $T\delta S$ is not a perfect differential. To
counterbalance the additional term one also needs to modify the expression
for $\delta S$. Here we take 
\begin{equation}
\delta S=2\pi k_{B}\left( \frac{1}{%
\lambdabar%
}\delta x+\frac{\beta }{mc}\delta p\right) ,  \label{E2}
\end{equation}%
with a dimensionless parameter $\beta $ introduced to the equation. We keep
the equation for the Unruh temperature (\ref{V3}) unchanged. In the limit $%
\alpha ,\beta \rightarrow 0$ we recover the equations (\ref{V1})-(\ref{V3}).
In \cite{santos2012} the options $\alpha =1$, $\beta =1$ and $\alpha =1$, $%
\beta =$ $p/mc$ were explored. Combining the equations (\ref{V3}), (\ref{E1}%
) and (\ref{E2}) leads easily to a corrected expression for the force\qquad 
\begin{equation}
F=ma+\left( \beta \frac{%
\lambdabar%
}{c}a-\alpha \frac{p}{m}\right) \frac{\delta p}{\delta x}=ma+F^{\text{cor}}.
\label{QF}
\end{equation}%
Our task is now to interpret the additional term $F^{\text{cor}}$ and test
which choices of $\alpha $ and $\beta $ are permissible. Keeping in mind
that the variations $\delta p$ and $\delta x$ are interpreted as the
uncertainties in a simultaneous measurement of $x$ and $p$ one can employ
the standard Robertson version of Heisenberg's uncertainty relation for a
simultaneous measurement of two noncommuting operators $A$ and $B$

\begin{equation}
\left( \Delta A\right) ^{2}\left( \Delta B\right) ^{2}\geq \frac{1}{4}%
\left\vert \left\langle \left[ A,B\right] \right\rangle \right\vert ^{2},
\label{HU}
\end{equation}%
for $A=x$, $B=p$ with $[x,p]=i\hbar $ to make estimations about the ratio $%
\delta p/\delta x$. In \cite{santos2012,ghosh2010planck} $\delta x\delta
p\geq \hbar /2$ was used at saturation point of the lower bound, i.e. it was
assumed that the test particle is in a coherent, possibly squeezed, state
and $\delta p$ was traded for $\hbar /(2\delta x)$. This leaves the
resulting expression with an unknown factor $\delta x^{-2}$. Furthermore it
was suggested in \cite{santos2012} that the classical limit is obtained by
the simultaneous limit $\hbar \rightarrow 0$, $\delta p\rightarrow 0$. This
gives indeed the classical expression, but the proposed prescription lacks
further justification. It addition, demanding $\delta p\rightarrow 0$ is
ambiguous as one might as well require the simultaneous limit $\hbar
\rightarrow 0$, $\delta x\rightarrow 0$. The question is why is the
classical limit not obtainable simply from $\hbar \rightarrow 0$? Here we go
a step further trying to achieve just that.

Our main assumption is that we take the test particle to be in a specific
state so that $\delta p/\delta x$ acquires a concrete value.

\subsection{Corrections from canonical coherent states and MOND dynamics}

Let us now assume at first the test particle to be in a standard canonical
coherent state for which we have the well known expressions, see e.g. \cite%
{gazeau2009coherent}, 
\begin{equation}
\delta x\delta p=\hbar /2,\qquad \ \ \ \ \ \ \text{with \ }\delta x=\sqrt{%
\frac{\hbar }{2m\omega }}\text{, ~\ }\delta p=\sqrt{\frac{m\omega \hbar }{2}.%
}
\end{equation}%
Using these equalities we can evaluate the ratio $\delta p/\delta x$ for
these states, such that the quantum corrected force (\ref{QF}) becomes%
\begin{equation}
F_{\text{coherent}}=ma+\left( \beta \frac{%
\lambdabar%
}{c}a-\alpha \frac{p}{m}\right) m\omega =ma+F_{\text{coherent}}^{\text{cor}}.
\label{Fco}
\end{equation}%
We have now various options for the choice of $\alpha $ and $\beta $. We may
take $\alpha \neq 0$, which suggests that the second term in (\ref{Fco})
becomes a pure quantum correction with $\alpha \sim \hbar $. For instance, $%
\alpha =\omega \hbar /mc^{2}$ is an admissible choice corresponding to a
quantum correction in the energy (\ref{E1}). Having introduced an additional
quantum correction it is not a surprise that we obtain also a quantum
correction in $F$.

Taking $\alpha =0$ the correction term $F_{\text{coherent}}^{\text{cor}}$
becomes a genuine quantum correction and the classical limit is simply
reached by taking $\hbar \rightarrow 0$. One might take $\beta =1$ in this
case, so that a classical correction in $\delta S$ has led us to a quantum
correction in $F$.

Finally one may wonder if one can reverse the setting of the previous
example and obtain a classical correction to $F$ from an additional quantum
term in $\delta S$. This is similar to Verlinde's original argument in which
also the $\hbar $ from the expression for $\delta S$ has cancelled the $%
\hbar $ appearing in the Unruh temperature. An example for such a classical
theory of modified Newtonian dynamics (MOND) was proposed in $1983$ by
Milgrom \cite{milgrom1983,milgrom2015road} for situations when the
gravitational force shows a marked departure from the conventional Newtonian
expectation at low acceleration. The MOND theory, or so it is called, is
typically applicable to scales of acceleration (less than the threshold
value of $a_{0}\thickapprox 1.2\times 10^{-10}ms^{-2}$) which are rather
small compared to what is observed in the solar system and perhaps relevant
towards explaining galactic scale phenomena \cite{smolin2017mond}. It has
been noted \cite{tortora2017} that the Milgrom scheme might be justified as
an alternative means to solve for the dark matter problem which is still to
find any experimental support, the prime reason being its rather poor
coupling with visible matter. A recent paper by Verlinde \cite%
{verlinde2017emergent} has sought to explore this issue using the standard
thermodynamical arguments as a basis.

A MOND theory has the force form given by a deformed acceleration \cite%
{gozzi2017newton}%
\begin{equation}
F_{\text{MOND}}=ma\mu (a_{0}/a),\qquad \text{with \ }\mu (a_{0}/a)=\frac{1}{%
1+a_{0}/a},  \label{MF}
\end{equation}%
where $a_{0}$ is the aforementioned small acceleration. It is, however, only
a phenomenological form but worthwhile to note that in place of the usual
expression of the acceleration as is implied by (\ref{1}) namely, $%
a=MGr^{-2} $, in MOND, a test particle which is at a distance $r$ from a
large mass $M$ is subject to the acceleration a given by $%
a^{2}/a_{0}=MGr^{-2}$, where $a\ll a_{0}$. Other variants of a modified
Newtonian equation have been proposed in the literature \cite{bhat2017m},
but we do not discuss them here.

Choosing now $\alpha =0$ and $\beta =-c(\omega 
\lambdabar%
)(1+a/a_{0})$ we obtain precisely the form of the MOND force (\ref{MF}) with
modified acceleration. Remarkably this means with a corrected entropy, just
taking the uncertainty of a particles position into account, we may
interpret the force in a MOND theory as a classically emerging entropic
force.

\subsection{Corrections from minimal value and minimal product coherent
states}

\label{corrmvmp}

Let us now assume our system to be in a noncommutative space on which the
canonical Heisenberg commutation relations are generalized to \cite%
{Kempf2,AFBB}%
\begin{equation}
\left[ x,p\right] =i\hbar \left( 1+\tau p^{2}\right) ,  \label{defHeisenberg}
\end{equation}%
with $\tau \in \mathbb{R}^{+}$ denoting the dimensionful, i.e. $\left[ \tau %
\right] =s^{2}/m^{2}$, noncommutative constant. Such a generalized
commutation relation arises as a particular case of an extended
q-deformation given by the commutator \cite{AFBB}

\begin{equation}
\lbrack x,p]=i\hbar q^{g(N)}+\frac{i\hbar }{4}(q^{2}-1)\left( \frac{x^{2}}{%
\delta ^{2}}+\frac{p^{2}}{\gamma ^{2}}\right) ,\quad \delta ,\gamma \in 
\mathbb{R}  \label{qdef}
\end{equation}%
where $g$ is some arbitrary function of the number operator $N$ defined as
the product of the creation and annihilation operator for the harmonic
oscillator. Taking $g(N)=0$ and parametrizing the deformation parameter $q$
in the form $q=e^{2\tau \gamma ^{2}}$, we found in the limit $\gamma
\rightarrow 0$, the $\tau $-corrected form (\ref{defHeisenberg}).

On a noncommutative space one needs to make an important distinction between
what we refer to as minimal factor coherent states (mfco) and minimal
product coherent states (mpco). The former are the states for which the
minimal value is reached for one of the factors in the uncertainty relation,
e.g. $\delta x$ in which case it is referred to as minimal length coherent
state (mlco). In contrast,the mpco-states is a state for which the entire
product in the uncertainty relation, e.g. $\delta x\delta p$, is minimized.
Assuming now the test particle to be in a mlco-state, we have 
\begin{equation}
\delta x_{\min }\delta p=\hbar ,\qquad \ \ \ \ \ \ \text{with \ }\delta
x_{\min }=\hbar \sqrt{\tau }\text{, ~\ }\delta p=\frac{1}{\sqrt{\tau }}.
\label{xpmin}
\end{equation}%
These values are easily obtained for (\ref{defHeisenberg}) with (\ref{HU}),
see the next section for the derivation. We will comment also in more detail
on the construction of meaningful explicit states that produce these values.
Using (\ref{xpmin}) in (\ref{QF}) the noncommutatively corrected force
becomes%
\begin{equation}
F_{\text{mlco}}=ma+\left( \beta \frac{%
\lambdabar%
}{c}a-\alpha \frac{p}{m}\right) \frac{1}{\hbar \tau }=ma+F_{\text{mlco}}^{%
\text{cor}}.
\end{equation}%
Since $\hbar $ as well as $\tau $ are very small, $\hbar \tau \ll 1$, the
correction term becomes very large, which does not make sense as we expect
only a small modification. However, by demanding that $\alpha \sim (\hbar
\tau )^{2}$ and $\beta \sim \hbar \tau $ this can be achieved. For instance, 
$\alpha =(\hbar \tau m\omega )^{2}$ and $\beta =\hbar \tau m\omega $ is an
admissible choice from a dimensional point of view. When using this option
the modifying terms proportional to $\delta p$ in (\ref{E1}) and (\ref{E2})
acquire a new interpretation. They are now noncommutative deformations that
give rise to an entropically emergent noncommutative space-time structure.

As is clear from the argument above, the deformed equation (\ref%
{defHeisenberg}) is only one of many possibilities obtainable from (\ref%
{qdef}) or other approaches. In \cite{AFBB} also higher order $\tau $%
-deformations were explored as for instance%
\begin{equation}
\left[ x,p\right] =i\hbar \left( 1+\tau p^{2}+\frac{1}{2}\tau
^{2}p^{4}\right) ,  \label{t2}
\end{equation}%
with $\delta x_{\min }=1.14698\hbar \sqrt{\tau }$, $\delta p=0.740664/\sqrt{%
\tau }$ going up to%
\begin{equation}
\left[ x,p\right] =i\hbar e^{\tau p^{2}},  \label{t3}
\end{equation}%
with $\delta x_{\min }=1.16582\hbar \sqrt{\tau }$, $\delta p=1/\sqrt{2\tau }$%
. It was noted in \cite{dorsch2012} that $1.16582\ldots $ can be expressed
as $\sqrt{e/2}$. This means for the higher deformations the corrected force
becomes 
\begin{equation}
F_{\text{mlco}}(\tau )=ma+\kappa _{\tau }F_{\text{mlco}}^{\text{cor}},
\end{equation}%
where $\tau $ in $\kappa _{\tau }$ indicates the order in the deformation.
So we have $\kappa _{1}=1$, $\kappa _{2}=0.6458$, \ldots , $\kappa _{\infty
}=e^{-1/2}=0.6065$.

One might suspect a different outcome in regard to the previous argument
when using mpco-states. Assuming still the emerging space-time structure to
be noncommutative with deformed uncertainty relation (\ref{defHeisenberg})
we obtain in this case 
\begin{equation}
\left( \delta x\delta p\right) _{\min }=\frac{2}{3}\hbar ,\qquad \ \ \ \ \ \ 
\text{with \ }\delta x=\frac{2}{\sqrt{3}}\hbar \sqrt{\tau }\text{, ~\ }%
\delta p=\frac{1}{\sqrt{3}\sqrt{\tau }}.  \label{xpminmin}
\end{equation}%
We note that indeed for these values the product $\delta x\delta p$ is
smaller and $\delta x$ is larger in (\ref{xpminmin}) when compared to (\ref%
{xpmin}). As it is less obvious how to derive these values, we will comment
in detail on the derivation in the next section. The correction term to the
force is in this case half the correction term $F_{\text{mpco}}=ma+F_{\text{%
mlco}}^{\text{cor}}/2$ which is a further reduction when compared to $\kappa
_{\infty }$. As the overall dependence on $\hbar $ and $\tau $ is unchanged
the general discussion and interpretation is the same as for the mlco-states
resulting from (\ref{defHeisenberg}).

\section{Minimal length and minimal value coherent states}

We will now provide the details on how the values for (\ref{xpmin}) and (\ref%
{xpminmin}) are obtained including a derivation of the associated explicit
wavefunctions. Since the discussion for the values in (\ref{xpminmin}) has
not been presented elsewhere before, we provide also a general discussion on
the appropriate method to be used.

We distinguish here two fundamental questions regarding the measurement of
an observable $A$ in any quantum mechanical system: a) what minimal value
can the variance $\left( \Delta A\right) ^{2}:=$ $\left\langle
A^{2}\right\rangle -\left\langle A\right\rangle ^{2}=\left\langle \hat{A}%
^{2}\right\rangle $ with $\hat{A}=A-\left\langle A\right\rangle $ or the
standard deviation $\Delta A$ take and b) what is the associated state $%
\left\vert \psi \right\rangle _{\min ,A}$ in $\left\langle A\right\rangle
:=\left\langle \psi \right\vert A\left\vert \psi \right\rangle $ that
realises that minimum? These questions are then naturally extended to
simultaneous measurements related to two or more operators. For two
operators $A$ and $B$ the questions a) and b) have now three variants, i.e.
what are the minimal values and corresponding states for $\Delta A$, $\Delta
B$ or the product $\Delta A\Delta B$ within the simultaneous measurement of $%
A$ and $B$? For three operators $A$, $B$ and $C$ this extends to seven
variants, i.e. what are the minimal values and corresponding states for $%
\Delta A$, $\Delta B$, $\Delta C$, $\Delta A\Delta B$, $\Delta A\Delta C$, $%
\Delta B\Delta C$ or $\Delta A\Delta B\Delta C$ within the simultaneous
measurement of the expectation values of all three operators $A$, $B$ and $C$%
? Naturally the states that realise these different possibilities are
usually non-identical.

A typical example for the measurement of two operators $A$ and $B$ are the
aforementioned position and momentum operators $x$ and $p$, respectively. To
measure the position of a particle in space is an example for three
operators corresponding to the coordinate components $x$, $y$ and $z$. This
measurement is of course trivial in conventional space, but becomes
nontrivial and interesting when one considers a noncommutative space in
which coordinate component operators do not commute, see e.g. \cite%
{AFLGBB,DFG} for concrete examples and \cite{Laurerev} for a recent review.

The question regarding the minimal values of $\Delta A$ and $\Delta B$ is
not challenging when the commutator on the right hand side of (\ref{HU}) is
a constant, as one can always achieve $\Delta A\rightarrow 0$ or $\Delta
B\rightarrow 0$ by taking $\Delta B\rightarrow \infty $ or $\Delta
A\rightarrow \infty $, respectively, and still respect the lower bound. This
is the standard scenario of Heisenberg's uncertainty relation in which one
must give up all the information about $A$ or $B$ to measure $B$ or $A$,
respectively, with absolute precision. However, when the resulting
commutator on the right hand side of the inequality involves operators, i.e.
when the lower bound becomes a function of $A$ and/or $B$ it is no longer
possible to carry out the limits in this trivial manner. Such a scenario
arises for theories formulated on certain noncommutative spaces as discussed
in section \ref{corrmvmp}. In that case one may assume that the minima are
reached for \textit{coherent states} that is at equality in (\ref{HU}). By
defining the function $f(\Delta A,\Delta B):=\Delta A\Delta B-\frac{1}{2}%
\left\vert \left\langle \left[ A,B\right] \right\rangle \right\vert $, the
critical values are simply obtained by simultaneously solving $f(\Delta
A,\Delta B)=0$ and $\partial _{\Delta B}f(\Delta A,\Delta B)=0$ for $\Delta
A $ or $\partial _{\Delta A}f(\Delta A,\Delta B)=0$ for $\Delta B$ from
which one can identify the minimum $\Delta A_{\min }$ or $\Delta B_{\min }$,
respectively, see e.g. \cite{AFBB}. We note that there is no reason to
expect that the product of the individual minimal values $\Delta A_{\min }$
and $\Delta B_{\min }$ is equal to the minimal value of the product $\left(
\Delta A\Delta B\right) _{\min }$. From this argument we do not obtain any
information about the states involved.

\subsection{The direct versus the analytic method}

Let us now see how to derive the associated minimizing states $\left\vert
\psi \right\rangle _{\min ,A}$, $\left\vert \psi \right\rangle _{\min ,B}$
and $\left\vert \psi \right\rangle _{\min ,AB}$ for which these minima are
reached. Following Jackiw \cite{jackiwmin} we recall the difference between
the \textit{direct} and \textit{analytic} method that determine the state $%
\left\vert \psi \right\rangle $ minimizing the uncertainty product. Making
the same assumption as above, namely that the minimum is reached for
equality, the direct method follows from a comparison of the Schwartz and
triangle inequality for $\left\vert \left\langle \hat{A}\hat{B}\right\rangle
\right\vert ^{2}$. It then consists of solving 
\begin{equation}
\left[ A-\alpha +\frac{\left\langle \left[ A,B\right] \right\rangle }{2b^{2}}%
(B-\beta )\right] \left\vert \psi \right\rangle =0  \label{DM}
\end{equation}%
for $\left\vert \psi \right\rangle $ involving the three free parameters $%
\alpha :=\left\langle A\right\rangle $, $\beta =:\left\langle B\right\rangle 
$ and $b^{2}:=\left\langle B^{2}\right\rangle -\left\langle B\right\rangle
^{2}$. Once the eigenvalue problem in (\ref{DM}) is solved, these parameters
are just computed in a self-consistent manner via their defining relations.
They may be used to convert the solutions into proper square integrable
functions and to minimize the desired quantities, that are either the
separate minimal values $\Delta A_{\min }$ , $\Delta B_{\min }$ or the
minimal product $\left( \Delta A\Delta B\right) _{\min }$. In the derivation
of (\ref{DM}) one makes two assumptions: First that the minimal state is
reached for the equality sign in (\ref{HU}) and second that the commutator $%
\left[ A,B\right] $ is a c-number rather than a q-number, that is a constant
and not an operator.

The analytic method on the other hand makes no assumptions about the right
hand side in the inequality (\ref{HU}). In that scheme one treats the left
hand side as a functional, minimizing it together with the supplementary
assumption that $\left\vert \psi \right\rangle $ is normalizable, i.e. one
solves 
\begin{equation}
\frac{\delta }{\delta \left\langle \psi \right\vert }\left[ \left(
\left\langle \psi \right\vert A^{2}\left\vert \psi \right\rangle
-\left\langle \psi \right\vert A\left\vert \psi \right\rangle ^{2}\right)
\left( \left\langle \psi \right\vert B^{2}\left\vert \psi \right\rangle
-\left\langle \psi \right\vert B\left\vert \psi \right\rangle ^{2}\right)
-m\left( \left\langle \psi \!\right. \left\vert \psi \right\rangle -1\right) %
\right] =0,  \label{func}
\end{equation}%
with Lagrange multiplier $m$, for $\left\vert \psi \right\rangle $. In a
straightforward manner this leads to the eigenvalue problem%
\begin{equation}
\left[ \frac{\left( A-\alpha \right) ^{2}}{a^{2}}+\frac{\left( B-\beta
\right) ^{2}}{b^{2}}\right] \left\vert \psi \right\rangle =2\left\vert \psi
\right\rangle .  \label{AM}
\end{equation}%
As no assumption is made about the right hand side in (\ref{HU}) an
additional parameter $a^{2}:=\left\langle A^{2}\right\rangle -\left\langle
A\right\rangle ^{2}$ enters the scheme when compared to the direct method.
By re-expressing the direct method as%
\begin{equation}
\left[ \frac{\left( A-\alpha \right) ^{2}}{a^{2}}+\frac{\left( B-\beta
\right) ^{2}}{b^{2}}\right] \left\vert \psi \right\rangle =2\frac{\left[ A,B%
\right] }{\left\langle \left[ A,B\right] \right\rangle }\left\vert \psi
\right\rangle ,
\end{equation}%
Jackiw \cite{jackiwmin} demonstrated that the two schemes coincide if and
only if $\left\vert \psi \right\rangle $ is an eigenstate of the commutator $%
\left[ A,B\right] $. Thus when this is not the case at least one of the
assumptions on which the direct method is based ceases to be valid.

As there are no obvious analogs to the Schwartz and triangle inequality for
three operators, it is not evident how to formulate the direct method for
that situation. However, it is straightforward to generalize the analytic
method to three observables, that is to minimize triple products of
variances $\left( \Delta A\right) ^{2}\left( \Delta B\right) ^{2}\left(
\Delta C\right) ^{2}$. Using the analogue to (\ref{func}), simply with an
additional factor $\left( \left\langle \psi \right\vert C^{2}\left\vert \psi
\right\rangle -\left\langle \psi \right\vert C\left\vert \psi \right\rangle
^{2}\right) $ on the first term, one easily derives 
\begin{equation}
\left[ \frac{\left( A-\alpha \right) ^{2}}{a^{2}}+\frac{\left( B-\beta
\right) ^{2}}{b^{2}}+\frac{\left( C-\gamma \right) ^{2}}{c^{2}}\right]
\left\vert \psi \right\rangle =3\left\vert \psi \right\rangle ,
\label{triple}
\end{equation}%
now with two additional free parameters $\gamma =\left\langle C\right\rangle 
$ and $c^{2}=\left\langle C^{2}\right\rangle -\left\langle C\right\rangle
^{2}$, see \cite{tripleWeigert} for an example computation and \cite%
{ma2017exp} for an experimental verification.

The advantage of the analytic over the direct method is that it is
applicable a) irrespective of the nature of $\left[ A,B\right] $, i.e.
resulting to a number or an operator b) even when the minimum is not reached
for equality in (\ref{AM})\ and c) to generalizations of uncertainty
relations involving more than two observables.

\subsection{Minimal value coherent states from direct method}

When one wishes to obtain more information about the wavefunctions $%
\left\vert \psi \right\rangle _{\min ,A}$, $\left\vert \psi \right\rangle
_{\min ,B}$ and $\left\vert \psi \right\rangle _{\min ,AB}$ via the direct
or analytic method one needs to specify a concrete representation for the
operators involved. For the algebra (\ref{defHeisenberg}) there are various
meaningful representations $\Pi _{(i)}$ that we label by $i$. For instance,
with regard to the standard inner product a non-Hermitian and a Hermitian
one are 
\begin{equation}
x_{(1)}=(1+\tau p_{s}^{2})x_{s},~p_{(1)}=p_{s},\quad \text{and\quad }%
x_{(2)}=x_{s},~p_{(2)}=\frac{1}{\tau }\tan \left( \sqrt{\tau }p_{s}\right)
,~~~  \label{22}
\end{equation}%
respectively. Here $x_{s}$ and $p_{s}$ are standard canonical variables
satisfying $[x_{s},p_{s}]=i\hbar $. Naturally for $\tau \rightarrow 0$ one
recovers the standard Heisenberg commutation relations. Models in terms of
these variables and further representations have been studied in more detail
in \cite{AFBB,HrepvsnHrep}. As demonstrated in \cite{HrepvsnHrep} the
corresponding physical quantities, namely expectation values for adjoint
operators, are representation independent and one may simply chose the most
suitable one for one's purpose.

Using now the non-Hermitian representation $\Pi _{(1)}$ in the equation for
the direct method (\ref{DM}) for the observables $A=x$ and $B=p$, we obtain
in momentum space simply a first order differential equation 
\begin{equation}
\left[ i\hbar \left( 1+\tau p_{s}^{2}\right) \partial _{p_{s}}+i\hbar \frac{%
1+\tau b^{2}+\tau \beta ^{2}}{2b^{2}}(p_{s}-\beta )-\alpha \right] \psi
_{d}(p_{s})=0,  \label{ddiif}
\end{equation}%
for the minimal state $\psi _{d}(p_{s})$. Setting the constant $\beta =0$,
equation (\ref{ddiif}) is easily solved to%
\begin{equation}
\psi _{d}(p_{s})=\left[ \frac{\sqrt{\tau }\Gamma \left( \frac{3}{2}+\frac{1}{%
2\tau b^{2}}\right) }{\sqrt{\pi }\Gamma \left( \frac{1}{2}+\frac{1}{2\tau
b^{2}}\right) }\right] ^{1/2}\left( 1+\tau p_{s}^{2}\right) ^{-\frac{1}{%
4\tau b^{2}}-\frac{1}{4}}\exp \left[ -\frac{i\alpha \arctan \left( p_{s}%
\sqrt{\tau }\right) }{\hbar \sqrt{\tau }}\right] .
\end{equation}%
The constant pre-factor is chosen so that these states are normalized with
regard to the quasi-Hermitian inner product%
\begin{equation}
\left\langle \psi \!\right. \left\vert \psi \right\rangle _{\rho
}:=\int\nolimits_{-\infty }^{\infty }\rho (p_{s})\psi ^{\ast }(p_{s})\psi
(p_{s})dp_{s}=1,  \label{product}
\end{equation}%
with metric operator $\rho =\left( 1+\tau p_{s}^{2}\right) ^{-1}$. As this
is by now well established in the literature we do not justify the choice of 
$\rho $ here any further, but instead refer the reader to \cite%
{Urubu,Benderrev,Alirev,siegl2012metric,bagarello2013self,FabFring1,krejvcivrik2015pseudospectra}
for the general formalism on how to construct meaningful metric operators
and how to define consistent inner products for non-Hermitian systems. A
well-known argument in \cite{Urubu} states that the metric becomes unique
when two observables with specific properties are specified, see also \cite%
{MGH}\footnote{%
In \cite{znojil2017} this argument was incorrectly attributed our previous
work \cite{AFBB}.}. Using the states $\psi _{d}$ in the expression $%
\left\langle .\right\rangle _{\rho }=\left\langle \psi _{d}\right\vert
.\left\vert \psi _{d}\right\rangle _{\rho }$, we then easily compute the
relevant expectation values%
\begin{equation}
\left\langle x\right\rangle _{\rho }=\alpha ,\quad \quad \left\langle
x^{2}\right\rangle _{\rho }=\alpha ^{2}+\frac{\hbar ^{2}(1+\tau b^{2})^{2}}{%
4b^{2}},\quad \quad \left\langle p\right\rangle _{\rho }=0,\quad \quad
\left\langle p^{2}\right\rangle _{\rho }=b^{2}.  \label{expect}
\end{equation}%
The values for $\left\langle x\right\rangle _{\rho }$, $\left\langle
p\right\rangle _{\rho }~$and $\left\langle p^{2}\right\rangle _{\rho }$ are
to be expected by definition from the formalism and the explicit
computations are just consistency checks. Minimizing $\left( \Delta x\right)
^{2}$ as a function of $b$ we find $b=1/\sqrt{\tau }$, such that the minimal
length becomes $\Delta x_{\min }=\hbar \sqrt{\tau }$, which coincides with
the findings reported in (\ref{xpmin}) and those in \cite{Kempf2}. For this
value of $b$ we have $\Delta p=1/\sqrt{\tau }$ so that the product $\Delta
x_{\min }\Delta p=\hbar $ does not saturate the lower bound. In the light of
the discussion in the previous section this was to be expected and in fact
the minimal value for $\Delta x\Delta p=\hbar /2(1+\tau b^{2})$, as well for 
$\Delta p=b$, would be obtained for $b=0$. However, while the corresponding
minimal length state $\left\vert \psi \right\rangle _{\min ,p}=$ $\psi
_{d}(b=1/\sqrt{\tau })$ is well-defined, $\psi _{d}(p_{s})$ is ill-defined
for $b=0$. So the direct method does not allow us to compute the product
coherent states $\left\vert \psi \right\rangle _{\min ,xp}$. Let us
therefore employ the analytical method to determine them.

\subsection{Minimal product coherent states from the analytical method}

For the representation $\Pi _{(1)}$ the eigenvalue equation for the
analytical method (\ref{AM}) in momentum space becomes the second order
differential equation 
\begin{equation}
\left[ -\frac{\hbar ^{2}\left( 1+\tau p_{s}^{2}\right) ^{2}}{a^{2}}\partial
_{p_{s}}^{2}-\frac{2\hbar (i\alpha +\hbar p_{s}\tau )\left( 1+\tau
p_{s}^{2}\right) }{a^{2}}\partial _{p_{s}}+\frac{\alpha ^{2}}{a^{2}}+\frac{%
(p_{s}-\beta )^{2}}{b^{2}}-2\right] \psi _{a}(p_{s})=0.
\end{equation}%
One may verify that $\psi _{d}(p_{s})$ does not satisfy this equation, which
is to be expected. Instead, setting $\beta =0$ this equation is solved in
terms of associated Legendre polynomials of $P_{\ell }^{m}(x)$ and Legendre
functions of the second kind $Q_{\ell }^{m}(x)$ 
\begin{equation}
\psi _{a}(p_{s})=\exp \left[ -\frac{i\alpha \arctan \left( p_{s}\sqrt{\tau }%
\right) }{\hbar \sqrt{\tau }}\right] \left[ c_{1}P_{\ell }^{m}(ip_{s}\sqrt{%
\tau })+c_{2}Q_{\ell }^{m}(ip_{s}\sqrt{\tau })\right] ,
\end{equation}%
with $\ell =\sqrt{4a^{2}+\hbar ^{2}\tau ^{2}b^{2}}/(2b\tau \hbar )-1/2$ and $%
m=a\sqrt{1+2\tau b^{2}}/(b\tau \hbar )$. Setting $\ell $ and $m$ to small
integer values we find the first meaningful solutions, in the sense of being
nonvanishing real numbers, for the parameters $a$ and $b$ for $\ell =1$ and $%
m=2$. For those values we have $P_{1}^{2}(ip_{s}\sqrt{\tau })=0$ and $%
Q_{1}^{2}(ip_{s}\sqrt{\tau })=2/(1+\tau p_{s}^{2})$. Suitably normalized
with regard to the inner product (\ref{product}) the minimizing solution
becomes%
\begin{equation}
\psi _{a}(p_{s})=\sqrt{\frac{8}{3\pi }}\frac{\tau ^{1/4}}{1+\tau p_{s}^{2}}%
\exp \left[ -\frac{i\alpha \arctan \left( p_{s}\sqrt{\tau }\right) }{\hbar 
\sqrt{\tau }}\right] .
\end{equation}%
Using this solution we then easily compute the expectation values%
\begin{equation}
\left\langle x\right\rangle _{\rho }=\alpha ,\quad \quad \left\langle
x^{2}\right\rangle _{\rho }=\alpha ^{2}+\frac{4\hbar ^{2}\tau }{3},\quad
\quad \left\langle p\right\rangle _{\rho }=0,\quad \quad \left\langle
p^{2}\right\rangle _{\rho }=\frac{1}{3\tau },
\end{equation}%
for $\left\langle .\right\rangle _{\rho }=\left\langle \psi _{a}\right\vert
.\left\vert \psi _{a}\right\rangle _{\rho }$. For these values the product
of uncertainties $\left( \Delta x\Delta p\right) _{\min }=2\hbar /3$ is
minimized by construction. Indeed this value is smaller than $\Delta x_{\min
}\Delta p$ obtained from the direct method by just minimizing $\Delta x.$ On
the other hand the uncertainty in $x$ computed from these states $\Delta x=2/%
\sqrt{3}\hbar \sqrt{\tau }$ is corrected by a factor $1.15$ and therefore
slightly larger than the one computed from the direct method or the
minimization of $f(\Delta x,\Delta p)$. Since the contributions from the two
observables to the minimal product is not the same, i.e. $\Delta x=2\hbar 
\sqrt{\tau /3}$ and $\Delta p=1/\sqrt{3\tau }$, the states $\psi _{a}(p_{s})$
are squeezed coherent states for all values of $\alpha $.

\section{Lagrangian and Hamiltonian formulation of the entropic force law}

Let us now investigate the modified force equation (\ref{QF}) further with a
special focus on the question of which choices for $\alpha $ and $\beta $
are meaningful. We observed in the previous section that an important
feature of the entropic force law is that it comes in two parts - one
corresponding to the inertial term which is the so-called Newtonian or
inertial mass $m_{I}$ times acceleration and the other due to quantum or
classical correction $F^{\text{cor}}$ which influences it. The entire
expression was derived purely on the basis of thermodynamical arguments and
the use of the uncertainty principle for particular states. The latter sets
a lower limit on the product of the uncertainties of position and canonical
momentum but a canonical momentum (defined as the partial derivative of the
Lagrangian with respect to the time-derivative of the generalized
coordinate) has no place in Newtonian theory which is concerned only with
the ordinary momentum given simply by mass times velocity. To keep things in
order it would be reasonable to identify the $p$ appearing in (\ref{QF})
with $\dot{x}$ whose rate of change is the acceleration $a=\ddot{x}$.

With this understanding we re-express the modified force equation(\ref{QF})
as

\begin{equation}
F=P\ddot{x}-Q\dot{x}=-\frac{\partial V}{\partial x},  \label{FF}
\end{equation}%
where the quantities $P$ and $Q$ stand for

\begin{equation}
P:=m+\beta \frac{\hbar }{mc^{2}}\frac{\delta p}{\delta x},\quad Q:=-\alpha 
\frac{\delta p}{\delta x}.
\end{equation}%
We also used in (\ref{FF}) the definition $F=-\partial V/\partial x$, where $%
V$ is a potential. It is straightforward to see that when taking

\begin{equation}
\mathcal{L}=\left( \frac{P}{2}\dot{x}^{2}-V\right) e^{-\frac{Q}{P}t}
\label{LL}
\end{equation}%
to be a Lagrangian, the corresponding Euler-Lagrange equation of motion

\begin{equation}
\frac{d}{dt}\left( \frac{\partial \mathcal{L}}{\partial \dot{x}}\right) =%
\frac{\partial \mathcal{L}}{\partial x}
\end{equation}%
is equivalent to (\ref{QF}) or (\ref{FF}) for time-independent $P$ and $Q$.
We notice that $\mathcal{L}$ depends explicitly on the damping factor%
\begin{equation}
\frac{Q}{P}=-\frac{\alpha \omega }{m\omega \frac{\delta x}{\delta p}+\beta 
\frac{\hbar \omega }{mc^{2}}},
\end{equation}%
which evidently needs to very small or vanishing altogether. We can use this
fact to constrain possible choices for $\alpha $ and $\beta $ even further.
For convenience we summarize previously discussed scenarios in table 1

\begin{center}
\begin{tabular}{|c|c|c|c|c|}
trial states & $\alpha $ & $\beta $ & $\delta x/\delta p$ & $\left\vert
Q/P\right\vert $ \\ \hline\hline
canonical coherent states & $1$ & $1$ & $\frac{1}{m\omega }$ & $\frac{\omega 
}{1+\hbar \omega /mc^{2}}\gg 1$ \\ \hline
canonical coherent states & $\frac{\hbar \omega }{mc^{2}}$ & $1$ & $\frac{1}{%
m\omega }$ & $\frac{\omega }{1+mc^{2}/\hbar \omega }\gg 1$ \\ \hline
canonical coherent states & $0$ & $1$ & $\frac{1}{m\omega }$ & $0$ \\ \hline
MOND & $0$ & $-c(\omega 
\lambdabar%
)(1+a/a_{0})$ & $\frac{1}{m\omega }$ & $0$ \\ \hline
mlco states & $\left( \hbar \tau m\omega \right) ^{2}$ & $\hbar \tau m\omega 
$ & $\hbar \tau /\kappa _{\tau }$ & $\frac{\hbar \tau m\omega ^{2}}{1/\kappa
_{\tau }+\hbar \omega /mc^{2}}\ll 1$ \\ \hline
mvco states & $\left( \hbar \tau m\omega \right) ^{2}$ & $\hbar \tau m\omega 
$ & $2\hbar \tau $ & $\frac{\hbar \tau m\omega ^{2}}{2+\hbar \omega /mc^{2}}%
\ll 1$%
\end{tabular}
\end{center}

\noindent {\small {Table 1: Choices for the dimensionless parameters $\alpha 
$, $\beta $ and different types of coherent states leading to admissible }}$%
{\small {(}}${\small {$\ll 1)$ and nonphysical $(\gg 1)$ damping factors.}}

\smallskip

The values in table 1 suggest that the framework discussed in \cite%
{santos2012}, based on a quantum correction for the energy is inconsistent.
However, two choices survive this simple test. First of all \textbf{any}
choice with $\alpha =0$ and arbitrary nonvanishing but dimensionally
acceptable $\beta $ is consistent. In particular that includes the values
leading to a classical MOND theory. Furthermore, the deformations of the
energy and entropy expressions corresponding to noncommutative deformations
also yield consistent damping factors.

Finally let us also comment on a Hamiltonian $\mathcal{H}$ that is readily
derived from the Langrangian $\mathcal{L}$ when taking the associated
canonical momentum $p$ (not to be confused with $\dot{x}$) to be

\begin{equation}
p=\frac{\partial \mathcal{L}}{\partial \dot{x}}=\dot{x}Pe^{-\frac{Q}{P}t}.~~
\end{equation}%
This leads us to

\begin{equation}
\mathcal{H}=\dot{x}\frac{\partial \mathcal{L}}{\partial \dot{x}}-L=\frac{%
p^{2}}{2P}e^{\frac{Q}{P}t}+Ve^{-\frac{Q}{P}t},
\end{equation}%
so that depending on the sign of $Q/P$ we obtain a damped kinetic term and
amplified potential term, or vice versa, unless $Q=0$.

Evidently the explicit time dependence in the coefficients provides a major
hindrance for the quantization of this system. However, one might follow for
instance recent work, using a modified Prelle-Singer approach, in which
explicit time-independent first integrals have been identified in different
parameter regimes for the damped linear harmonic oscillator problem in order
to facilitate the quantization procedure \cite{chandrasekar2007}.

\section{Conclusions}

We have re-examined Verlinde's argument on the derivation of the entropic
law of gravitation by taking into account a possible modification in the
form of adding a term linearly dependent on the momentum uncertainty $\delta
p$ in the energy (\ref{E1}) and/or the entropy (\ref{E2}) uncertainty. These
terms involve two free parameters $\alpha $ and $\beta $ that are only
constraint by dimensional arguments. We then computed the general correction
term for the entropic force resulting as a consequences of this modification
for various choices for the parameters. The uncertainty ratio $\delta
p/\delta x$, on which these terms depend, was computed exactly for some
concrete coherent states with underlying standard and generalized Heisenberg
uncertainty relation. In addition, we derive explicit expressions for the
normalizable wavefunctions associated to minimal length and minimal value
coherent states. Furthermore, we found that the Lagrangian and Hamiltonian
associated to the entropic force are explicitly time-dependent in an
exponential form. Demanding for obvious physical reasons that this
exponential is damped provided us with a further criterion to select
possible values for the parameters $\alpha $ and $\beta $.

On a standard space with conventional commutation relations for $x$ and $p$
we found consistent correction terms for $\alpha =0$ accompanied by any
choice for $\beta $. The latter can be chosen to implement classical or
quantum corrections in the entropy. Remarkably a specific version of the
latter option allows for the emergence of a \underline{classical} MOND
theory (\ref{MF}). Keeping $\alpha =0$ other choices for $\beta $ can lead
to classical as well as quantum corrections. Any scenario with $\alpha \neq
0 $ and $\beta \neq 0$ leads to exponentially growing Langrangians and
Hamiltonians and therefore appear to be inconsistent from a physical
perspective. This includes the scenario advocated in \cite{santos2012}.

However, when considering the equations in a noncommutative setting with
deformed canonical commutation relations (\ref{defHeisenberg}), (\ref{t2}), (%
\ref{t3}) dimensional consistency together with the requirement that the
classical and noncommutative limit can be carried out by $\hbar \rightarrow 0
$ and $\tau \rightarrow 0$, respectively, leads to consist solutions with a
very mild damping or amplifying factor in the Langrangian. One may say that
this setting leads to an emergent theory of noncommutative space-time.

There are many interesting issues left to investigate, as for instance to
clarify the link to a relativistic theory and to extend the analysis to
higher dimensions, using (\ref{triple}), to name only two.

\bigskip \noindent \textbf{Acknowledgments:} One of us (BB) thanks Siddharth
Seetharaman for useful comments.


\newif\ifabfull\abfulltrue

\end{document}